\documentstyle[pre,aps,epsfig,multicol]{revtex}

\newcommand{\mb}[1]{\mbox{\boldmath $#1$}}

\begin{document}

\draft

\title{Hierarchical noise in large systems of independent agents}

\author{Claus Wilke and Thomas Martinetz}

\address{Institut f\"ur Neuroinformatik, Ruhr-Universit\"at Bochum,
  D-44780 Bochum, Germany}

\date{Submitted: ; Printed: \today}

\maketitle

\begin{abstract}
A generalization of the coherent-noise models [M. E. J. Newman and
K. Sneppen, Phys. Rev. E{\bf54}, 6226 (1996)] is presented where the
agents in the model are subjected to a multitude of stresses,
generated in a hierarchy of different contexts. The hierarchy is
realized as a Cayley-tree. Two different ways of stress propagation in
the tree are considered. In both cases, coherence arises in
large subsystems of the tree. Clear similarities between
the behavior of the tree model and of the coherent-noise model can be
observed. For one of the two methods of stress propagation, the
behavior of the tree
model can be  approximated very well by an ensemble of coherent-noise
models, where the sizes $k$ of the systems in the ensemble scale as $k^{-2}$.  
The results are found to be independent of the tree's structure for a
large class of reasonable choices.
Additionally, it is found that power-law distributed lifetimes of
agents arise even under the complete absence of correlations
between the stresses the agents feel.

\end{abstract}

%\newpage

%===================================================================

\begin{multicols}{2}

\section{Introduction}

It has recently been shown  that in large systems of independent
``agents'', the interplay of two different types of noise  can lead
to power-law distributed 
quantities, like life-times of the agents, or sizes of reorganization
events~\cite{NewmanSneppen96}. One of the two noises, usually referred
to as stress, has to 
act coherently on all agents, while the other one, usually referred to
as mutation or reloading, has to act individually
on each agent, and on a much longer time scale. Mechanisms of this
kind have been called ``coherent-noise'' mechanisms. Models
incorporating coherent-noise mechanisms have
been put forward to explain effects seen in earthquakes, rice-piles, or
biological evolution and
extinction~\cite{NewmanSneppen96,Newman96,Newman98,WilkeMartinetz97}. 

 In most applications, however, it is hard to
justify a single stress imposed on
the whole system at once.  In~\cite{Newman96},
the stress was identified with global influences on the
biosphere, as in the case of extraterrestrial
impacts~\cite{Alvarez87}. Nevertheless, there are more 
reasons for species to go extinct than impacts. Often, species'
extinction is a local phenomenon~\cite{Raup93}. For example, species 
living in a small territory regularly die out because 
of the invasion of a new species, able to exploit their
ecological niche more effectively. A similar argument applies to the
situation of earthquakes. In~\cite{NewmanSneppen96} the stress has
been interpreted 
as background noise with long wavelength, generated by some distant external
source. Nonetheless, in a large fault system, we would expect
background noise to be present also locally, and probably on smaller
and larger scales at the same time~\cite{Kagan94}. 

The aim of the present paper is to advance a model that, while incorporating
the basic ideas of coherent-noise systems, can deal with more
complex situations by considering stress on different scales. A short
account of this work has already been given
elsewhere~\cite{WilkeAltmeyerMartinetz98}. There, only regular trees
(see below) have been treated.

\section{Agents in a hierarchical context}

It is an observation from every-day life, as well as from many
physical systems~\cite{Nicolis86}, that very often objects
or agents are embedded into a hierarchy of different contexts,
all having influences on them. 
Mathematically, this idea can be described with the concept of
ultrametricity~\cite{RammalToulouseVirasoro86}, which means there
exists a distance $d(\cdot,\cdot)$ such that the triangle inequality
$d(A_1,A_3)\leq \max\{d(A_1,A_2), d(A_2,A_3)\}$ holds for any three
agents $A_1$, $A_2$, $A_3$. Geometrically, an ultrametric space can be
conceived of as a Cayley-tree. In the following, vertices in the tree
connected to exactly one other 
vertex will be called leaves, and vertices connected to two or more
other vertices will be called nodes. In this paper, the nodes of the
tree stand for different contexts, and the agents are placed at the tree's
leaves. 

We can formulate a 
generalization of the original coherent-noise model by incorporating
the above ideas. Our system
consists of $N$ agents, each represented by a real number
(threshold) $x_i$,
or, in the general case, a vector $\mb{x}_i$. Furthermore, we choose a
tree with $N_{\rm n}$ nodes and $N_{\rm l}=N$ leaves, which means
$N_{\rm v}=N_{\rm n}+N_{\rm l}$ vertices in total. The tree will
be kept fixed 
throughout the simulation. At every leaf we put exactly one
agent. For every node $j$ of the tree we choose a stress distribution
with probability density function (PDF)
$p_j(x)$. Additionally, we also choose a stress distribution for
every leaf of the tree, so that we have a stress distribution for
every vertex of the tree.  The stress distributions  
at the leaves allow to simulate extremely localized influences
acting on only a single agent. The total number of stresses in the system
is therefore $N_{\rm stress}=N_{\rm v}$. 

The course of the simulation runs as follows. At the beginning, the
agents are initialized with random thresholds drawn from a distribution
$p_{\rm thresh}(x)$. Then, in every time step, three actions are
performed: i) from each of the $N_{\rm stress}$ stress distributions, a stress
$\eta_j$ is chosen at random. ii) for every agent $i$, from all the ${S_i}$
stress values $\eta_1^{(i)},\dots,\eta_{S_i}^{(i)}$ above the agent in the
tree, a stress 
$\eta_i^{\rm eff}$ is calculated according to some function $\cal
A$:
\begin{equation}
  \eta_i^{\rm eff}={\cal
  A}\Big(\eta_1^{(i)},\dots,\eta_{S_i}^{(i)}\Big)\,.
\end{equation}
If $\eta_i^{\rm eff}\geq x_i$, agent $i$ is removed and replaced
by a new one with a threshold chosen at random from $p_{\rm
  thresh}(x)$. iii) finally, every agent has a 
small probability $f$ to get a new threshold, again from the
distribution $p_{\rm thresh}(x)$.   
Action~iii) represents the mutation or reloading mentioned in the
introduction. 

There are a number of reasonable choices for the function $\cal A$. In
this paper, we will mainly study the ``maximum rule'', which reads
\begin{equation}\label{eq:funcAmax}
  {\cal A}\Big(\eta_1^{(i)},\dots,\eta_{S_i}^{(i)}\Big)=
    \max \Big\{ \eta_1^{(i)},\dots,\eta_{S_i}^{(i)}\Big\}\,.
\end{equation}
Another natural choice is to sum up the stresses,
i.e., to use
\begin{equation}\label{eq:funcAsum}
 {\cal A}\Big(\eta_1^{(i)},\dots,\eta_{S_i}^{(j)}\Big)=
    \sum_{j=1}^{S_i}\eta_j^{(i)}.
\end{equation}
This alternative, which we will call ``sum rule'', will also be
discussed in this paper.

\section{The effective stress distribution}
\label{sec:eff_dist}
The effective stress an agent feels can be calculated exactly in the
case of the maximum rule, Eq.~(\ref{eq:funcAmax}).
The agent is subjected to the stress distribution at its leaf
and to the stress distributions 
at the nodes above it. Let there be $S-1$ nodes above 
the leaf of an agent. Then the $S$ stress values having influence on
this agent are 
$S$ random variables $X_1, \dots, X_S$ with PDF's
$p_1(x), \dots, p_S(x)$. To obtain the effective stress distribution,
we have to calculate the PDF $p_{\max}(x)$ of the 
random variable $X_{\max}=\max\{X_1, \dots, X_S\}$, i.e.,
\begin{equation}\label{eq:pmaxdef}
  p_{\max}(x)\,dx=P\Big(x\leq\max\{X_1, \dots, X_S\}<x+dx\Big)\,.
\end{equation}
Note that 
\begin{equation}\label{eq:max_calc}
  P\Big(\max\{X_1, \dots, X_S\}\leq x\Big)=\!\!\!\prod_i^S P(X_i<x)\,.
\end{equation}
The derivative of Eq.~(\ref{eq:max_calc}) with respect to $x$ yields
\begin{equation}\label{eq:pmax} 
  p_{\max}(x)=
  \sum_{i=1}^S\, p_i(x) \!\!\! \prod_{j=1, j\neq i}^S\!\! P(x > X_j) 
\end{equation} 
Eq.~(\ref{eq:pmax}) is the exact expression for the effective stress
on an agent in the case of the maximum rule Eq.~(\ref{eq:funcAmax}).
A simple calculation shows that the right-hand side of
Eq.~(\ref{eq:pmax}) is dominated by the slowest decaying stress distribution. 
We say that a 
distribution $p_i(x)$ decays slower than another distribution $p_j(x)$ if
there exists a $x_0$ such that
\begin{equation}\label{eq:fall_off_slower}
  p_i(x)>p_j(x) \qquad \mbox{for all $x>x_0$.}
\end{equation}
For a set of reasonable stress distributions it is always possible to identify
the distribution $p_0(x)$  that is falling off slowest according to this definition.
Hence, we find for the PDF of the effective stress on an agent
\begin{equation}\label{eq:maxlimit}
  p_{\max}(x) \sim p_0(x) \quad\mbox{for $x\rightarrow\infty$} \,.
\end{equation}

A similar statement cannot easily be proved for the sum rule.
However, in special cases the necessary
calculations can be done. Consider, for example, the case of
exponential stress distributions
$p_i(x)=\exp(-x/\sigma_i)/\sigma_i$. In this case, we find
\begin{equation}\label{eq:exp_stresslimit}
  p_{\rm sum}(x)\sim  \frac{1}{\sigma_{\max}}
  \exp(-\frac{x}{\sigma_{\max}})  \quad \mbox{for $x\rightarrow\infty$,}
\end{equation}
where $\sigma_{\max}=\max\{\sigma_1,\sigma_2,\dots,\sigma_n\}$. A
similar result can be found in the case of stress distributions with
power-law tails~\cite{WilkeAltmeyerMartinetz98}. It seems that in
most of the 
reasonable cases, the sum of the stresses will be 
dominated by a single distribution in the limit of large stresses,
as in the situation of the maximum of the
stresses.

\section{Regular trees} 
\label{sec:regtrees}

\begin{figure}
\narrowtext
\centerline{
        \epsfig{file={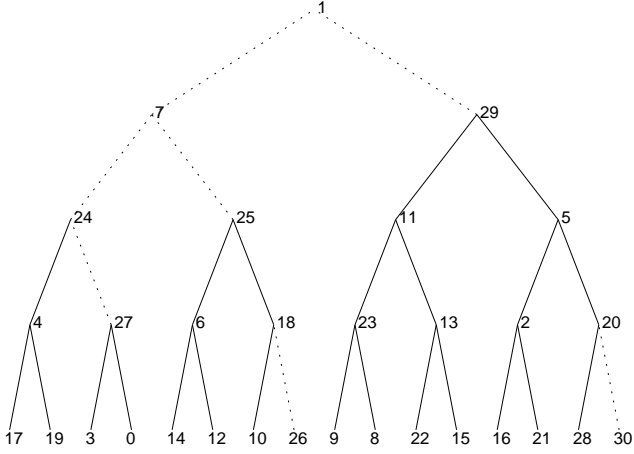}, height=\columnwidth, angle=270}
}
\caption{The breakdown of a regular tree with n=2 and l=4 into
  independent subsystems. The solid lines connect agents with the same
  rank, the dashed lines connect agents with different ranks. In this
  example, we have two subsystems of size 1 (ranks 26 and 30), two
  of size 2 (ranks 24 and 27), one of size 3
  (rank 25), and one of size 7 (rank 29). 
\label{fig:ranking}}
\end{figure}

In this section we are interested in trees which are constructed as
follows. We begin with a single leaf and convert it into a node by
connecting to it $n$ new leaves. Then we repeat this procedure for
every new leaf. We stop the construction when we have reached a depth
of $l$ iterations. Trees generated in this way are called 
regular trees~\cite{OlemskoiKiselev98}. The tree displayed in
Fig.~\ref{fig:ranking} is a regular tree with $n=2$ and $l=4$.

 The number of leaves in a regular tree is 
\begin{equation}\label{eq:Nleavesreg}
  N_{\rm l}= n^l\,,
\end{equation}
and the number of vertices is
\begin{equation}\label{eq:Nvertreg}
  N_{\rm v}= \sum_{i=0}^{l-1}n^i\,.
\end{equation}
Therefore, we have $N=n^l$ agents in such a tree, and 
$N_{\rm v}$ stress values have to be generated in every time step.  

We saw in Sec.~\ref{sec:eff_dist} that every agent feels
effectively a single stress distribution in the limit of large
stresses (Eqns.~(\ref{eq:maxlimit}), (\ref{eq:exp_stresslimit})).
Since for coherent-noise systems, large stresses give the main
contribution to the systems' behavior, the stress distributions that
are falling of very slowly dominate large parts of the tree.
Hence, the tree breaks down into subsystems that are to some extent
decoupled from each other.
For regular trees, it is relatively easy to
study the average distribution of the subsystems' sizes analytically. 
We assume
for any two stress distributions $p_i(x), p_j(x)$ in the tree we can
identify one of the two that is falling off slower,  according
to Eq.~(\ref{eq:fall_off_slower}). This is not a severe restriction,
as we have noted in Sec.~\ref{sec:eff_dist}. Additionally, we restrict
ourselves to situations in which $p_i(x)$ and $p_j(x)$ are equally
likely to fall off slower than the respective other.  Under these conditions,
we can rank all stress distributions in a tree, assigning rank 1 to
the one that is falling off fastest, and assigning correspondingly higher ranks
to the ones that are falling off slower. 
This makes the calculation of the
subsystems' sizes relatively easy. For every single agent, we have to
identify the corresponding highest rank placed above it in the tree
(which we will call the rank
of the agent). Then, we simply have to count the number of agents
with the same rank. 
This procedure is illustrated in Fig.~\ref{fig:ranking}.

We expect the mean distribution of subsystems' sizes to have sharp
peaks whenever the size of a complete subtree is reached, because the
probability for a single rank to be higher than all others further
down the tree should be larger than the probability for a complicated
arrangement of ranks to produce a subsystem of a certain
size. The size $k$ of a subsystem is the number of leaves in that
subsystem. The functional dependency of the peaks at size
$k$ is calculated as follows. The expected frequency 
$f(k)$ of independent subtrees of depth $b$,
corresponding to a subsystem of size $k=n^b$, can be written as the
number of such subtrees in the whole system, $N_{\rm sub}(n^b)$, times
the probability that 
any of these subtrees will be independent of the rest, $P_{\rm
  indep}(n^b)$. Hence we write
\begin{equation}
  f(n^b)=N_{\rm sub}(n^b)P_{\rm indep}(n^b)\,.
\end{equation}
The number of subtrees of size $n^b$ is
\begin{equation}\label{eq:Nsub}
 N_{\rm sub}(n^b)=n^{l-b}\,. 
\end{equation}
A subtree is independent of the rest if the rank at its root is higher
than all other ranks in the subtree and at the nodes above the
subtree. The probability $P_{\rm indep}(n^b)$ is therefore the reciprocal
of the number of vertices in the subtree plus the number of vertices
above the subtree, hence
\begin{equation}\label{eq:Pindep}
  P_{\rm indep}(n^b)=\Big(l-b+\sum_{i=0}^{b}n^i\Big)^{-1}\,.
\end{equation}
If we increase $b$ by one, we get
$N_{\rm sub}(n^{b+1})=n^{l-b-1}=N_{\rm sub}(n^b)/n$. With 
slightly more effort, we find also
\begin{eqnarray}
  P_{\rm indep}(n^{b+1})&=&\Big(l-b-1+\sum_{i=0}^{b+1}n^i\Big)^{-1}
                                 \nonumber\\ 
 &=&\Big(l-b+n\sum_{i=0}^{b}n^i\Big)^{-1}
 \approx \frac{1}{n}P_{\rm indep}(n^b)\,.
\end{eqnarray}
Therefore, we can write
\begin{equation}\label{eq:regtreescaling}
 f(nk)\approx \frac{N_{\rm sub}(k)}{n}
       \frac{P_{\rm indep}(k)}{n} =n^{-2}f(k)\,,
\end{equation}
which implies $f(k)\sim k^{-2}$.

This result is interesting. The frequency of subsystems of size $k$
scales as $k^{-2}$, independent of the parameter $n$ which
characterizes the structure of the tree. 

We have tested these predictions by measuring the frequency $f(k)$ in
computer experiments. Our simulations are set up as follows. We choose
a tree with $N_{\rm v}$ vertices in total. For several thousand
times, we assign the integers from 1 to $N_{\rm v}$ randomly
to the vertices of the tree. The integers stand for the rank of the
stress-distributions at the vertices. For every single realization of this
process, we determine the sizes of the subsystems the tree breaks down
into, and compute a histogram of the sizes' frequencies. Finally, we
calculate the average over all histograms.

\begin{figure}
\narrowtext
\centerline{
        \epsfig{file={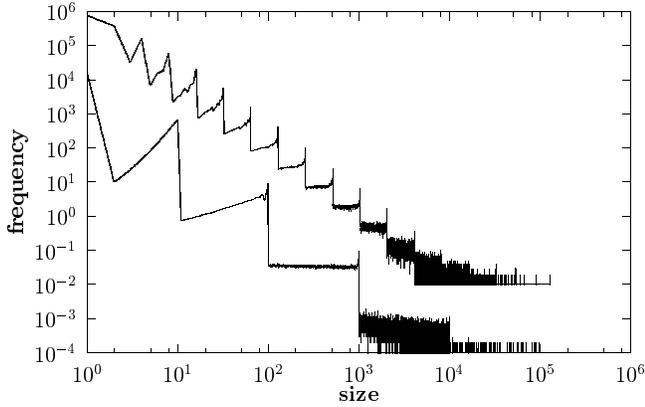}, width=\columnwidth}
}
\caption{The expected frequency of subsystems of size $k$ decreases as
  a sawtooth function following approximately a power-law with
  exponent $-2$. The upper curve stems from a tree with $l=17$ and
  $n=2$. It has been rescaled by a factor of 100 so as not to overlap
  with the lower curve. The lower curve stems from a tree with $l=5$
  and $n=10$.
  Quantities are plotted in arbitrary units.
\label{fig:parts}}
\end{figure}

Fig.~\ref{fig:parts} shows the results of such measurements for two
different trees with 10 000 histograms each. We can see clear peaks
at powers of $n$, which correspond to complete subtrees. We also find
the heights of the peaks to decrease as $k^{-2}$, in agreement with
Eq.~(\ref{eq:regtreescaling}).

\section{Random trees}
\label{sec:randtrees}

The regular trees treated in the previous section can be easily
generalized to a broader class of trees, which we will call ``random
trees''. Only a small change in the construction algorithm is
necessary. To construct 
a regular tree, in every iteration step we connect $n$ new leaves to
every leaf of the previous step. The straightforward generalization of
this procedure is
to choose a random number of new leaves for every leaf of the previous
construction step. To avoid confusion with the parameter $n$, we will
call this random number 
$n_{\rm rand}$. The random variable $n_{\rm rand}$ will take value $i$
with probability $p_i$, i.e., $P(n_{\rm rand}=i)=p_i$,
$i=0,1,2,\dots$, $\sum_i p_i=1$. We denote the mean of $n_{\rm rand}$ 
by $m:=\langle n_{\rm rand} \rangle$ and the variance 
by $\sigma^2$. Moreover, we assume $m>1$ in all cases
considered in this paper. In the limit
$\sigma^2\rightarrow 0$, the random trees reduce to regular trees with
$n=m$. 

The construction of a random tree as prescribed above is a branching
process with $l$ generations. From the theory of branching
processes~\cite{Harris63} ,
we know that for large $l$ the number of leaves in the tree will be
\begin{equation}\label{eq:Nleavesrand}
  N_{\rm l}=Wm^l\,,
\end{equation}
where $W$ is a random variable with mean $\langle W\rangle =1$. The
factor $W$ takes 
into account fluctuations that happen 
at the beginning of the tree's construction. Correspondingly, for
the total
number of vertices in the tree we use the approximation
\begin{equation}\label{eq:Nvertrand}
  N_{\rm v}\approx W\sum_{i=0}^{l}m^i\,.
\end{equation}
The above two equations are the generalizations of
Eqs.~(\ref{eq:Nleavesreg}), (\ref{eq:Nvertreg}) for random trees.

As in the case of regular trees, we are interested in the quantity
$f(k)$, the expected frequency with which independent subsystems of size
$k$ occur. In  the previous section we made the 
assumption that the main contributions to $f(k)$
come from complete subtrees. The comparison with numerical data showed
that this assumption leads to a good understanding of the structure of
$f(k)$. Consequently, in the case of random trees we also assume that
we can concentrate on complete subtrees.

The number of subtrees of size $k$ in a large tree is on average the
size of the tree (which is the number of leaves in the tree) divided
by $k$. Hence we have 
\begin{equation}\label{eq:Nsubrand}
  N_{\rm sub}(k)=\frac{N_{\rm l}}{k}\,,
\end{equation}
which is equivalent to Eq.~(\ref{eq:Nsub}) for regular trees.

The probability
for a subtree of size $k$ to be dominated by a single stress
distribution is  one
over the total number of vertices in the subtree. The number of
vertices is asymptotically the same as the number of
leaves. This can be seen from Eqns.~(\ref{eq:Nleavesrand}) and
(\ref{eq:Nvertrand}). The leading term in the number of vertices in a
random tree Eq.~(\ref{eq:Nvertrand}) is exactly the expression for the
number of leaves in the same tree Eq~(\ref{eq:Nleavesrand}). Hence we
have
\begin{equation}
  P_{\rm indep}(k)\sim \frac{1}{k}\,.
\end{equation}
We combine this result with Eq.~(\ref{eq:Nsubrand}) and obtain 
\begin{equation}
  f(k)\sim\frac{1}{k^2}\,.
\end{equation}
As in the case of regular trees, the frequency of independent subtrees
of size $k$ scales as $k^{-2}$, independent of the details of the
tree. With a little effort, it is also possible to calculate the
constant of proportionality. We find
\begin{equation}\label{eq:fk_rand_approx}
  f(k) = \frac{\alpha N_{\rm l}}{k^2}\frac{m-1}{m}\,,
\end{equation}
with
\begin{equation}
  \alpha=\left[\sum_{k=1}^{N_{\rm l}}
      \frac{1}{l+1-\log_{m}k +(k-1)\frac{m}{m-1}}\right]^{-1}\,.
\end{equation}

\begin{figure}
\narrowtext
\centerline{
        \epsfig{file={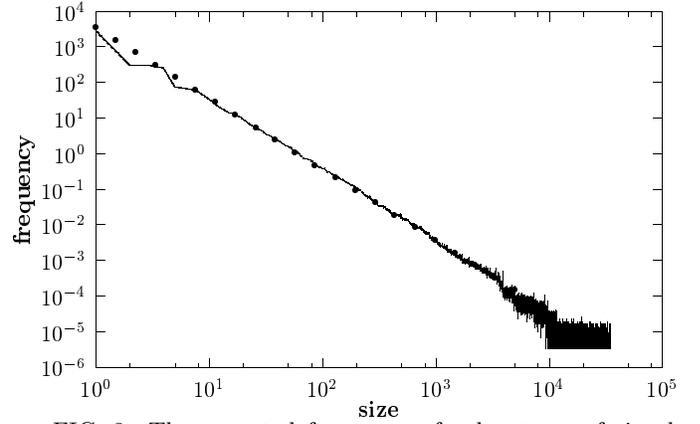}, width=\columnwidth}
}
\caption{The expected frequency of subsystems of size $k$ in a random
  tree with $m=1.993$ %nmax=5, Nl=34325
  and $l=13$. The dotted line is the
  approximation Eq.~(\ref{eq:fk_rand_approx}).
  Quantities are plotted in arbitrary units.
\label{fig:rand_tree}}
\end{figure}

Eq.~(\ref{eq:fk_rand_approx}) is in good agreement 
with measurements from computer experiments. We have done simulations
with several different probability distributions for $n_{\rm rand}$,
such as uniform [$p_0 = c/(n_{max}+c)$; $c\geq 0$; $p_i = 1/(n_{max}+c)$ for
$1\leq i\leq n_{\max}$; $p_i = 0$ for $n_{\max}<i$], geometric
series [$p_i = bc^{i-1}$ for $i\geq 1$; $b,c>0$; $b\leq 1-c$;
$p_0=1-\sum_{j=1}^\infty p_j$], or gaussian [$p_i\sim \exp
(-(i-b)^2/c)$; here $b$ and $c$ are not mean and
variance, because we use only discrete values of the gaussian
probability density function]. In all cases, we find
Eq.~(\ref{eq:fk_rand_approx}) to approximate well the measured frequency
$f(k)$. An example is shown in
Fig.~\ref{fig:rand_tree}. Deviations from the straight line can be
seen for very small $k$ and for very large $k$. In these two limiting
cases, the assumptions of the above approximations are no longer
valid. Consider first the case of a very small $k$. This corresponds
to $k\approx m$, because we always assume $m\ll N_{\rm v}$ (otherwise,
the tree would have roughly a depth of 1, which would not be very
interesting). If $k$ is close to $m$, the number of subtrees of size
$k$ depends strongly on the exact form of the probability distribution of
$n_{\rm rand}$, and Eq.~(\ref{eq:Nsubrand}) is no longer valid. Since,
as seen above, 
the main contribution to $f(k)$ comes from complete subtrees,
the distribution of $n_{\rm rand}$ then has effects on 
$f(k)$. For example, in a situation where 
$P(n_{\rm rand}=m)=0$, there should be a clear dip in $f(k)$ at $k=m$.

Consider now the case of a very large $k$. Again
Eq.~(\ref{eq:Nsubrand}) is no longer valid. This time because there are so
few subtrees of size $k$. Hence, the exact structure of the tree comes
into play. For example, a tree with $N_{\rm v}=10^5$ containing a
subtree with $k=6\times 10^4$ will not contain another subtree with
$k=5\times 10^4$. Therefore, in this situation there should be a clear
peak at $k=6\times 10^4$.

\section{Simulation Results}

As the main result of Sections~\ref{sec:regtrees}
and~\ref{sec:randtrees}, we found the distribution of independent
subsystems of size $k$ in the tree to be proportional to
$k^{-2}$. Therefore, in the limit of large stress values $\eta$, we
expect the tree model to behave like an ensemble 
of coherent-noise models whose sizes scale as $k^{-2}$. 

\begin{figure}
\narrowtext
\centerline{
        \epsfig{file={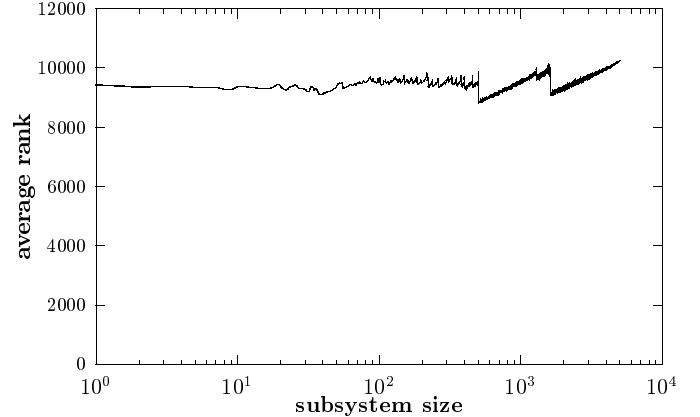}, width=\columnwidth}
}
\caption{\label{fig:stresses}The average rank of the subsystems in a
  random tree.
  Quantities plotted are dimensionless.
}
\end{figure}

When constructing the ensemble approximation of a certain tree, we
have to choose the right stress distribution for every coherent-noise
model in the ensemble. In principle, this can be a complicated
task. However, we have found that a very simple approach instead works
sufficiently well in many cases. It can be motivated with
Fig.~\ref{fig:stresses}. There,  we have
recorded the average ranks of the subsystems in a tree. Interestingly,
the average
rank varies only very little with the subsystem's size $k$. Therefore,
in a further approximation, we assume that all the
stress-distributions that dominate a subsystem have the same rank,
i.e., they are all the same (we use  a single \emph{stress
  distribution}, but the 
\emph{stress values} are still chosen independently for all systems in
the ensemble).

Our numerical simulations show the similarity between the tree model
and the ensemble. We begin with results for the maximum rule.

\subsection {The distribution of event sizes}

Like in previous work~\cite{SneppenNewman97,WilkeAltmeyerMartinetz97},
an ``event'' is the reorganization of agents because of stress
in a single time step. The size of an event is the total number of
agents hit by the stress.

\begin{figure}
\narrowtext
\centerline{
        \epsfig{file={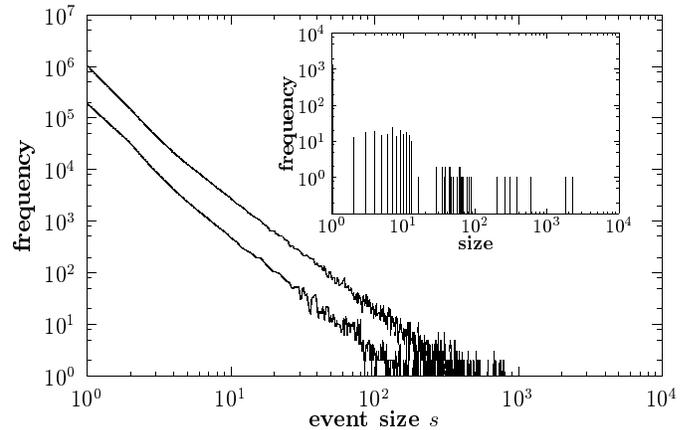}, width=\columnwidth}
}
\caption{\label{fig:exdist1}The event size distribution in a random
  tree (lower curve) and in the corresponding ensemble of
  coherent-noise systems (upper curve). The inset shows the
  distribution of the tree's subsystems.
  Quantities are plotted in arbitrary units. 
}
\end{figure}

The event sizes of a typical simulation with maximum rule are recorded in
Fig.~\ref{fig:exdist1}. The lower curve shows the distribution of event
sizes of a tree model, the upper curve shows the same distribution
of the corresponding ensemble of coherent-noise systems. The heights
of the curves reflect the total number of events we recorded for each
model and have no special meaning. The sizes of the systems in the
ensemble are exactly the ones we obtained for the sizes of the tree's
subsystems while doing the ranking procedure described in
Sec.~\ref{sec:regtrees}. The distribution of these
sizes is shown in the inset of Fig.~\ref{fig:exdist1}.

The tree used in Fig.~\ref{fig:exdist1} is a random tree with 14213
vertices and 12163 leaves. Hence, both the tree model
and the ensemble contain 12163 agents in total. The stress
distributions used in the tree are exponentials
$\exp(-x/\sigma_i)/\sigma_i$, with different values $\sigma_i$ between
0.03 and 0.06. The stress distribution used in the ensemble is an
exponential with $\sigma=0.06$.

As we can see in Fig.~\ref{fig:exdist1}, the tree model and the
ensemble behave very similar with regard to event sizes. In both
cases, we find approximately a power-law decrease. A power-law fit
gives an exponent of $2.3\pm 0.15$ for the tree model, and of $2.2\pm0.15$
for the ensemble. Note the clear difference between the exponent in
these two systems and  the exponent in a single coherent-noise model
with exponential stress. There, the exponent is
$1.85\pm 0.03$~\cite{SneppenNewman97}.

\begin{figure}
\narrowtext
\centerline{
        \epsfig{file={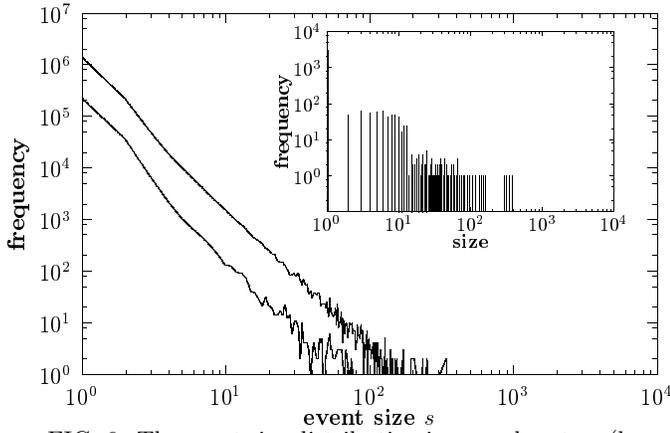}, width=\columnwidth}
}
\caption{\label{fig:exdist2}The event size distribution in a random
  tree (lower curve) and in the corresponding ensemble of
  coherent-noise systems (upper curve). The structure of the tree is
  the same as in 
  Fig~\ref{fig:exdist1}, but the stress distributions at the vertices
  are different. The inset shows again the distribution of subsystems 
  obtained from the ranking procedure.
  Quantities are plotted in arbitrary units.
}
\end{figure}

The event-size distribution depends
strongly on the distribution of the subsystems in the tree. In
Fig.~\ref{fig:exdist2}, we have used the same tree structure and the
same stress distributions as in Fig.~\ref{fig:exdist1}, but the stress
distributions have been assigned to different vertices. As a result,
in this case the tree model has a lack of large
subsystems, as can be seen in the inset of Fig.~\ref{fig:exdist2}. 
Consequently, large events appear less frequently, and the distribution
is significantly steeper than in Fig.~\ref{fig:exdist1} (now we have
an exponent of 
$2.9\pm0.2$ for the tree model and an exponent of $2.8\pm0.2$ for the
ensemble). 

\begin{figure}
\narrowtext
\centerline{
        \epsfig{file={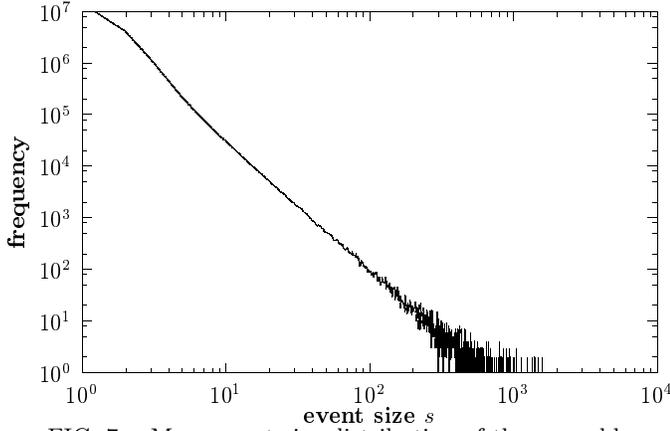}, width=\columnwidth}
}
\caption{\label{fig:SumHisto} Mean event size distribution of the
  ensemble approximation to a tree with 10000 leaves. The average was
  taken over 60 randomly generated ensembles. 
  Quantities are plotted in arbitrary units.}
\end{figure}

It would be interesting to average over all possible assignments of
the stress distributions to the different vertices in order to gain a better
understanding of a typical event size distribution in a large
tree. However, we are not able to reach such a result due to the
enormous amount of computing power that is needed. The
simulation of the full 
tree as in Figs.~\ref{fig:exdist1} or \ref{fig:exdist2} takes in fact about
150 hours of computing time on a UltraSPARC 2 with 168 MHz. On the
other hand, the simulation of the 
ensemble approximation takes only 6 hours on the same
system. Therefore, we can do the corresponding calculations for the
ensemble approximation. Of course we cannot average over
all possible configurations, but we can average over a reasonably
large random sample. We have generated 60 ensemble approximations
of a tree with 10000 leaves. The event size distributions we found
were all very similar. In Fig.~\ref{fig:SumHisto} we
display the average event size distribution we obtained. The
distribution has a power-law tail with exponent $2.5\pm 0.05$.

\subsection {Aftershocks}

\begin{figure}
\narrowtext
\centerline{
        \epsfig{file={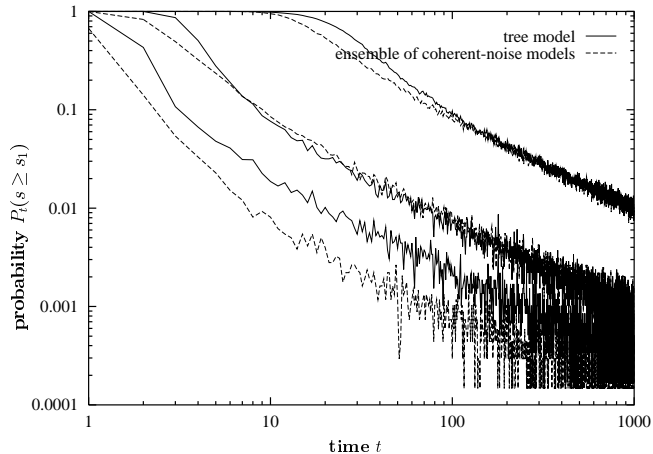}, width=\columnwidth}
}
\caption{\label{fig:aftershocks}The probability $P_t(s\geq s_1)$. The
  solid lines stem from the tree model, the dashed lines stem from the
  corresponding ensemble. From bottom to top, we have $s_1=0.05$,
  $s_1=0.02$ and $s_1=0.0025$, where $s_1$ is measured in units of the
  maximal system size. For the upper two curves, the results for the
  tree model are very close to the ones for the ensemble. The
  discrepancies in the lower curve are explained in detail in the
  text. Quantities plotted are dimensionless.}
\end{figure}

Coherent-noise models display
aftershocks~\cite{NewmanSneppen96,WilkeAltmeyerMartinetz97}, i.e., an
increased number of large events can be observed in the aftermath of a
very large event.
Consequently, we study the decay pattern of the aftershocks in the
tree model and in the ensemble. We will restrict ourselves 
to the case of events in the aftermath of an initial infinite
event. We follow closely the ideas and methods developed
in~\cite{WilkeAltmeyerMartinetz97}. Fig.~\ref{fig:aftershocks} shows
the change of the probability $P_t(s\geq s_1)$ with time. $P_t(s\geq
s_1)$ is the probability to find an event larger than some constant
$s_1$ at time $t$ after an initial infinite event. For both the tree
model and the ensemble, the probability $P_t(s\geq s_1)$ decreases with
time, indicating aftershocks. However, we do not observe a clear power-law
decrease, normally visible in the case of coherent-noise
models~\cite{WilkeAltmeyerMartinetz97}. 

As in the case of event-sizes, we find a close similarity between the
tree model and the ensemble. Let us first focus on the upper two
curves in Fig.~\ref{fig:aftershocks}, which correspond
to $s_1=0.02$ and $s_1=0.0025$ (here, $s_1$ is measured in units
of the number of agents in the tree, which was 12163 in this
case). For large $t$, the curves for the tree 
model and for the ensemble lie on top of each other, indicating the
same decay pattern for long-time correlations. Only for
small times there are some deviations between the two models. The tree
model produces more aftershocks shortly after the infinite
event. This observation has its origin in the fact that the two models
converge in the limit of large stresses, but the number of moderate
stresses produced by the tree model is significantly larger than the
one produed by the ensemble. At short times after a very large event,
already moderate stresses can trigger large events, thus increasing
the number of events seen in the tree model as compared to the ensemble.

\begin{figure}
\narrowtext
\centerline{
        \epsfig{file={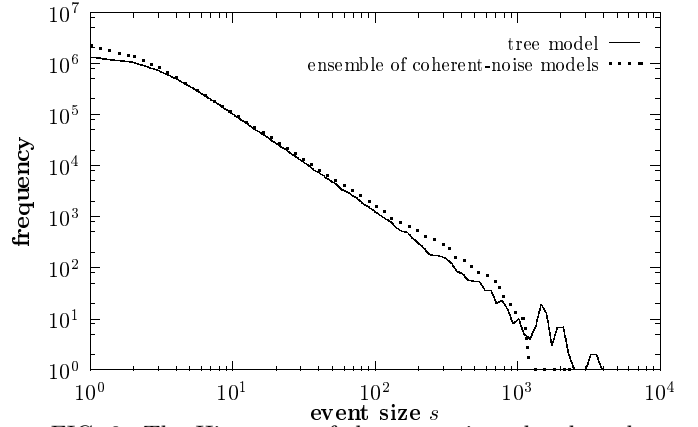}, width=\columnwidth}
}
\caption{\label{fig:afterevents}The Histogram of the event sizes that
  have been produced in the simulations for
  Fig.~\ref{fig:aftershocks}. Note that we recorded events only up to
  1000 time steps after the infinite event. Therefore, the exponent
  of the power-law is different from the one in
  Fig.~\ref{fig:exdist1}.
  Quantities are plotted in arbitrary units.}
\end{figure}

For large values of $s_1$, the similarity between the two models
seems to disappear. The curves in Fig.~\ref{fig:aftershocks} corresponding to
$s_1=0.05$  do not lie on top of each other. The curve for the tree
model is shifted upwards by about a factor of three. This discrepancy
for large $s_1$ can be understood from
Fig.~\ref{fig:afterevents}. There, we display the frequency distribution
of the events that have been produced during the simulations for
Fig.~\ref{fig:aftershocks}. The results for the tree model and for the
ensemble are very similar. However, at an event-size of about 1000,
the frequency distribution for the ensemble falls off rather quickly,
whereas the frequency distribution for the tree has an additional peak
at about 1400. It is this peak that causes the shift of the
probability $P_t(s\geq s_1)$ in the tree model for large $s_1$.

The peak in the tree model arises  because from time to time a very
large stress will be generated at the root of the tree, causing an
event of the order of the tree's size. In the ensemble, on the other
hand, events larger than the largest subsystem are extremely
unlikely.

\subsection {The distribution of lifetimes}
\label{sec:lifetimes}

The
lifetime of an agent is the time an agent remains in the system
without being hit by stress. In the original
coherent-noise model, the agents' lifetimes are 
distributed as a power-law with exponent
$2-1/\alpha$~\cite{SneppenNewman97}. The quantity 
$\alpha$ depends on the stress distribution, and it is related to the
mean-field exponent $\tau$  of the event size distribution 
by $\tau=1+\alpha$. For exponential stress, e.g., we have
$\alpha=1$. Hence, in this case the lifetimes $L$ are distributed as
$L^{-1}$. 

\begin{figure}
\narrowtext
\centerline{
        \epsfig{file={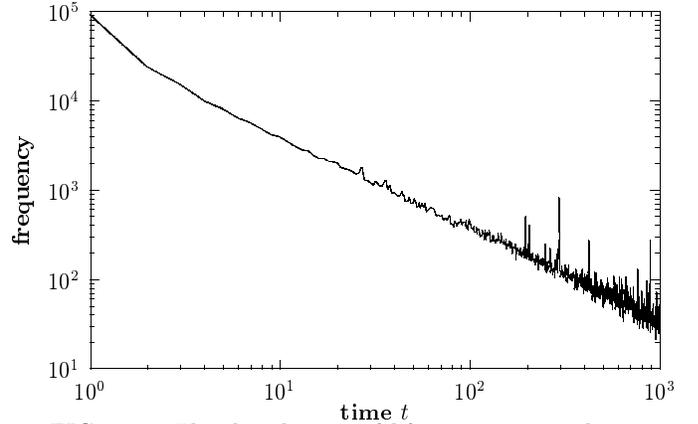}, width=\columnwidth}
}
\caption{\label{fig:lifetimes} The distribution of lifetimes in a
  random tree with exponential stresses
  only. We find a power-law with an exponent of -1.02. This is the
  same result as in a  
  coherent-noise model with exponential stress.
  Quantities are plotted in arbitrary units.}
\end{figure}

The distribution of lifetimes in a
coherent-noise model does not change
if the stress is imposed on each agent independently, instead of being
imposed on all agents coherently. This is different to the case of
event sizes or aftershocks. It can be seen as follows. The 
derivation of the lifetime distribution in~\cite{SneppenNewman97}
makes use of the time-averaged distribution of the agents' thresholds,
which remains the same whether or not the stress is imposed
coherently. The only 
further assumptions that enter the calculation are assumptions about
the form of $p_{\rm stress}(x)$ and $p_{\rm thresh}(x)$, but no
assumptions about the coherence of stresses are made. Therefore, the
distribution of lifetimes in a coherent-noise model and
in a large ensemble of degenerate coherent-noise models with
size 1 is the same, provided the stress distributions and the
threshold distributions are the same.
Consequently, if the stress-distributions in the tree have
all the same $\alpha$ (e.g., are all exponentials), the distribution
of the agents' lifetimes should be similar to the one in a
coherent-noise model with $\tau=1+\alpha$. This can be seen in
Fig.~\ref{fig:lifetimes}. The distribution of lifetimes in a random
tree with exponentially distributed stresses is similar to the one in
a coherent-noise model with exponential stress (compare, e.g., 
Fig.~\ref{fig:lifetimes} with Fig.~5 in~\cite{SneppenNewman97}).

\subsection{Trees with sum rule}
\begin{figure}
\narrowtext
\centerline{
        \epsfig{file={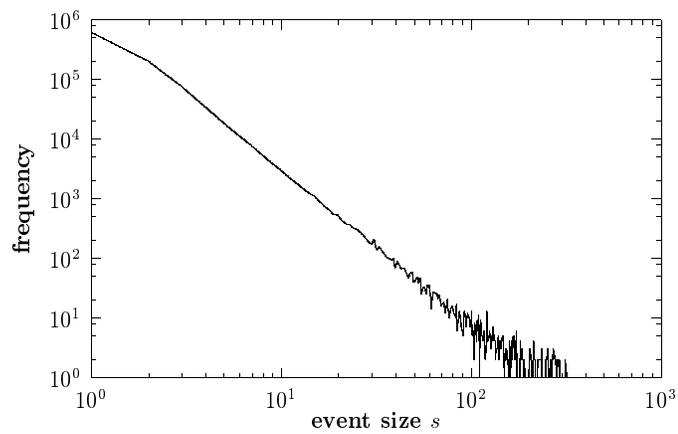}, width=\columnwidth}
}
\caption{\label{fig:sum}The distribution of event sizes in a tree
  model where the stresses are summed up. The tree (including the
  stress distributions) is exactly the same as in
  Fig.~\ref{fig:exdist1}.
  Quantities are plotted in arbitrary units.}
\end{figure}

In the previous paragraphs, we studied simulations with the maximum rule.
Here, we will present some results from simulations with the sum rule.
On the first glance, one would expect that the tree
model behaves the same whether we choose Eq.~(\ref{eq:funcAmax}) or
Eq.~(\ref{eq:funcAsum}) for calculating the effective stress on the
agents, at least for exponential stress, because of
Eq.~(\ref{eq:exp_stresslimit}). However, this is  not exactly the case. 
In Fig.~\ref{fig:sum}, we display the distribution of event sizes in a
simulation where stresses are 
summed up. The tree used in this
simulation is exactly the same we used in the simulation of
Fig.~\ref{fig:exdist1}. This allows an easy comparison between the two
choices for $\cal A$. Note that all stress distributions are
exponentials, which implies that Eq.~(\ref{eq:exp_stresslimit}) holds.
We observe the emergence of a power-law decrease, similar to the
situation with the maximum rule. However, 
the resulting distribution is
slightly steeper than in Fig.~\ref{fig:exdist1}, with an exponent of
$2.6\pm0.1$. 
This steeper distribution shows that the conception of a tree being
equivalent to an ensemble of coherent-noise models is less accurate
when stresses are summed up. Second-order effects arise because all
stress distributions contribute to the overall system's behavior at
all times (which is in contrast to the case when we use the maximum
of the stresses). Consequently, the agents feel the stress less coherently,
resulting in a smaller number of large events. 

\section{Conclusions}

Coherent-noise models have been proposed by Newman and Sneppen to
explain the occurence of power-laws in a number of natural
systems. The underlying mechanism is remarkably simple and robust.
However, the coherent stress
necessary to make these models work is an impediment to their
application, since in most systems coherence is not
present \emph{a priori}, and local phenomena are important. In this paper, we
were able to show
that in hierarchical contexts, coherence can arise naturally in
large subsystems. In the tree models we presented, the system breaks
down into a number of subsystems, 
each of them having a high degree of coherence and being largely
independent of the rest. Interestingly, the number of
subsystems of size $k$ decreases as $k^{-2}$ for a large class of
different trees. The emergence of coherent subsystems is closely
connected to the domination of some stress distributions by others. We
should always observe this phenomenon if the function $\cal
A$ is proportional to a single stress distribution in
the limit of large stresses.

We made also an interesting observation about the agents lifetimes. We
found the distribution of lifetimes to be the same in the tree model and in
coherent-noise models, as long as the stress distributions in both
models have the same functional dependency. Furthermore, from the
arguments given 
in Section~\ref{sec:lifetimes} we can deduce an even more general
statement. In any system where agents under the influence
of stress are modeled as in coherent-noise systems, the
distribution of the agents' lifetimes will be a power-law, even if
there is no correlation between stresses different agents feel. 
This is a new explanation for the appearance of power-law
distributed life-times or waiting-times in non-equilibrium systems
valid under extremely weak conditions. 

Further work extending the tree model presented here could
address appearance and disappearance of agents. If we consider, for
example, the case of biological evolution and extinction, the
biodiversity is constantly changing, with the main tendency of
exponential growth throughout the past 1000 million
years~\cite{Benton95} (this trend, however, has changed dramatically nowadays,
because of ever increasing human activity~\cite{Myers97}). "Real"
extinction and speciation could be incorporated into the tree model by
removing from the tree the agents hit by stress, as it has
been done already in the case of coherent-noise
models~\cite{WilkeMartinetz97}. Related to this, one
could consider 
trees changing their structure. Up to now, we studied only
fixed trees, mainly for reasons of simplicity. 
Another extension could be the consideration of vector stresses,
as it has been done by Sneppen and Newman for the original
coherent-noise model~\cite{SneppenNewman97}, inspired by a similar
generalization of the Bak-Sneppen model~\cite{BoettcherPaczuski96}.

\begin{acknowledgments}
We thank Thomas Daube for useful comments at an early stage of the
manuscript and Xavier Gabaix for pointing us to a
simplified calculation of the effective stress distribution. We also
would like to thank our unknown referee for his useful suggestions.
\end{acknowledgments}

\newpage

\end{multicols}

\end{document}